\documentstyle[12pt]{article}
\begin{document}
\baselineskip 24pt
\bibliographystyle{unstr}
\vbox{\vspace{6mm}}
\begin{center}
{\large \bf \baselineskip=24pt Is There a  Classical Analog of a Quantum
Time-Translation Machine?}
 \\[7mm] Lev Vaidman\\
{\it School of Physics and Astronomy
\\Raymond and Beverly Sackler Faculty of Exact Sciences
\\ Tel--Aviv University,
\ \ Tel--Aviv \ \ 69978 \ \ ISRAEL}\\[5mm]

\vspace{12mm}

ABSTRACT
\end{center}
\vspace{2mm}

In a recent article [D. Suter, Phys. Rev. {\bf A 51}, 45 (1995)]   Suter
has claimed to present an optical implementation
of the quantum time-translation machine  which ``shows all the features
that the general concept predicts and also allows, besides the quantum
mechanical, a classical description.'' It is  argued that the experiment
proposed and performed by Suter does not have the features of the   quantum
time-translation machine and that the latter has
no classical analog.

\break

A quantum time machine, suggested by Aharonov et al. \cite {AAPV} and
elaborated by Vaidman \cite {V}, is not a realistic device for a time
travel.  It is a gedanken procedure which, it seems, have no chance for
practical implementation in a near future. Moreover, even on the level of
a gedanken experiment, the machine usually fails to operate. Only very
rarely it succeeds to operate, but if it does, it achieves what, as far as
we know, no other machine can.

 Let us spell out what this quantum time machine does when it succeeds to
work.  Let the time of the operation of the machine to be $T$ and the time
evolution parameter of the machine to be $T'$. If we put inside the time
machine
any system (which fulfills some general requirements of energy spectrum
boundness) then, at
the end of the operation, the system will evolve to the state in which it
would have been after the undisturbed evolution of the time $T'$ (instead of
$T$).  The
important property  is that we do not have to know which  system  was put
inside and what  was its initial state. Our machine has an indicator
saying that the time-translation was accomplished successfully, and this
without ``looking'' on the system inside.

There are two well known classical devices which perform this task. In
fact, they achieve even more, since they work always, and there are less
restrictions
on what system can be put inside. The first device is
a fast rocket which makes a round trip, and the second is a heavy massive
shell which is placed around the system for a period of time. However, all
classical time machines can change the effective evolution time from $T$ to
$T'$ with the restriction $0<T'<T$, i.e., a classical time machine can only
slow down the time evolution.  Contrary to this, in the quantum time machine
the parameter $T'$  can have an arbitrary value. For $T'>T$ it speeds up
the time evolution and for $T'<0$ it effectively changes the direction of
the time flow. Neither of these effects can be achieved classically.

Suter \cite {SUT} has claimed to perform an experimental realization of the
quantum time-translation machine using a classical Mach-Zehnder
interferometer, and he concluded that the time-translation effect of the
proposed quantum time machine is ``not specific to quantum mechanics, but
is well known in classical field theory.''  The experimental setup of
Suter, however, does not fall even close to the definition of the time
machine. In his setup we know what is the system and what is its initial state.
What he shows is
that if we send a single mode of a radiation field through a birefringent
retardation device which yields different retardations for two orthogonal
polarizations, then placing the pre-selection polarization filter and
the post-selection polarization filter will lead to a much larger effect
than  for any of the pre-selected only polarizations. So, it might seem
like speeding up the time evolution, but this procedure fails all
tests of universality.  Different modes of radiation field  speed up
differently, an arbitrary wave packet is usually
distorted, and for other systems (other particles) the device is not
supposed to work at all.

So the first basic requirement that the time machine has to work for
various systems is not fulfilled from the beginning. And it cannot be
easily modified since the ``external'' variable
(which is supposed to be a part of the time-machine) is the property of the
system itself -- the polarization of the radiation field.
 The next necessary
requirement, that it works for a large class of the initial states of the
system, cannot be fulfilled too.  Indeed, he considers a superposition of only
two time evolutions.  This superposition  can be identical to a longer
evolution
for a particular state, but not for a large class of states. As it has been
shown \cite {AAPV,V} a superposition of a large number of time evolutions is
necessary
for this purpose.

We have shown that the experiment of Suter is not an
implementation of the quantum
 time machine. Still, it can be interpreted as a {\em weak measurement}
\cite{AAV,AV}.
The experiment  of Suter with a birefringent retardation device can be
considered as a weak measurement of a polarization operator. In fact, this
is a variation of the experiments which were proposed \cite {KV} and
performed \cite {RSH} previously. In the first proposal, a birefringent
prism caused an unusually large deflection of a well localized beam with  pre-
and
post-selected  polarization. And, in the experiment which was successfully
performed
\cite {RSH}, a birefringent plate was used instead.
 The ``weakness condition'' of these two experiments
follows from the localization of the beam (which was sent  through a narrow
slit).
The ``weak'' regime of the experiment of Suter is
achieved by taking the retardation small,  $\delta << \pi$.

 All these three
experiments can be explained both as quantum experiments on the ensembles of
photons or as ``classical'' experiments with electromagnetic waves.
We believe, however, that the correct way to consider the weak measurement
effect is
as a genuine  effect of quantum mechanics. First, because  it has no
analog in classical {\em mechanics}. Second, because in general, it has also no
explanation in terms of classical {\em waves}. As we shall explain below, it is
an accidental fact that all
performed weak measurement experiments can be explained in terms of  classical
waves.

 According to the conceptual procedure  of weak measurements, we start with an
ensemble of  systems prepared in the same  state, we perform (weak)
measurements with the measuring devices (one labeled device for each system),
we make post-selection measurements on
all systems, we discard all the result of the measuring devices which
correspond to the systems which did not yield the appropriate result in the
post-selection measurement, and finally we make statistical analysis
(calculating the avarage) of the readings of the remaining measuring devices.
All actually
performed weak measurements had the property that the pointer variables of
the measuring devices
were the variables of the systems themselves. This property made the weak
measurement much
easier to perform, but it is in no way a necessary property of weak
measurements. This is, however, a necessary property
for having explanation in terms of classical waves. If the measuring device
and the measured system are indeed different systems, than the weak
measurement effect follows from the quantum correlations between these two
systems which have no classical analog.
Even if the variable of the measuring device is a variable of the system itself
(which makes   the post-selection
much easier to perform), the classical explanation is not necessarily granted.
 The
experiments which were performed used  lasers, i.e., classical sources
of light. However,  nobody doubts that the same
results would be obtained if  sources of single photons were used instead, and
these
experiments have no classical description.

We  also question the claim of Suter, that
 the   interference effect of
classical  waves, which appears  in certain  weak measurements, is well known.
 It is
 well known, indeed, that pre- and post-selection might lead to unusual
interference effects. However, the point of the weak measurements is not just
that
the final superposition is very different from its components, but that
it  has a
particular form described by a very general and simple formula \cite{AV}. The
final
distribution  has (almost) the same form as the initial
one, and  its center is shifted to a well defined value, given
by  the {\em weak value} of the measured
quantity $A$, $A_w \equiv \langle \Psi_2 |A|\Psi_1 \rangle
/ \langle \Psi_2 |\Psi_1 \rangle$, where $|\Psi_1\rangle$ is the
pre-selected state and  $|\Psi_2\rangle$ is the
post-selected state. We do not know
that this remarkable feature was ever seen before.

We want to mention  here that Suter, together with R. Ernst and M. Ernst,
performed in the past
another experiment which they called ``An
experimental realization of a quantum time-translation machine'' \cite
{SEE}. In this experiment a very different system was used: the effect was
demonstrated on the heteronuclear coupling between two nuclear spins. But the
experimental setup was also applicable only to a specific system and only for a
certain state. Therefore, the same criticism is applicable and, therefore, one
should not
call it an implementation of the time-translation machine. One might, however,
consider it as another  successful  weak measurement,  the weak measurement of
a nuclear spin component.

\end{document}